# Effect of tungsten on vacancy behaviors in Ta-W alloys from first-principles

Yini Lv[a], Kaige Hu[b], Shulong Wen[c], Min Pan[c*], Zheng Huang[a], Zelin Cao[a], Yong Zhao[b]

[a] *School of Physical Science Technology, Southwest Jiaotong University, Chengdu,Sichuan, 610031, China*

[b] *School of Physics and Optoelectronic Engineering, Guangdong University of Technology, Guangzhou, Guangdong, 510006, ChinaChina.*

[c] *Superconductivity and New Energy R&D Center, Key Laboratory of Advanced Technology of Materials (Ministry of Education), Southwest JiaoTong University, Chengdu, Sichuan, 610031, China.*

## Abstract

Alloying elements play an important role in the design of plasma facing materials with good comprehensive properties. Based on first-principles calculations, the stability of alloying element W and its interaction with vacancy defects in Ta-W alloys are studied. The results show that W tends to distribute dispersedly in Ta lattice, and is not likely to form precipitation even with the coexistence of vacancy. The aggregation behaviors of W and vacancy can be affected by their concentration competition. The increase of W atoms has a negative effect on the vacancy clustering, as well as delays the vacancy nucleation process, which is favorable to the recovery of point defects. Our results are in consistent with the defect evolution observed in irradiation experiments in Ta-W alloys. Our calculations suggest that Ta is a potential repairing element that can be doped into Ta-based materials to improve their radiation resistance.

## 1. Introduction

Magnetic confinement tokamak is the most likely solution to realize controlled thermonuclear fusion. However, the working environment of high temperature and high neutron energy irradiation (~14MeV) will cause damage to plasma facing materials (PFMs), reducing their service performance. Among the first wall candidate materials, tungsten (W) has been widely studied in recent years because of its high melting point, high thermal conductivity, low sputtering yield, no chemical etching with hydrogen (H) and low H retention [1]. However, the main disadvantages such as low temperature brittleness, recrystallization brittleness and irradiation damage hinder the application of W. Recent studies have shown that tantalum (Ta) has a higher flux threshold for ion-induced surface nanostructures than W [2], which will reduce the damage risk to material integrity and contamination of reactor plasma. Due to its high density, high temperature resistance, corrosion resistance, good plasticity at low temperature and moderate elastic modulus [3-6], Ta has been considered to be a potential material for high temperature applications such as nuclear engineering, aerospace and other fields [7-11]. Ta alloys also have strong resistance to corrosive substances and high temperature radiation [12-14]. Therefore, Ta has been a candidate PFM in future fusion reactors recently. Adding alloying elements is a significant way to improve the radiation resistance of pure materials [15]. Ta-W alloys have attracted great interests because of their excellent properties at high temperature [16-22]. Microstructure evolution of Ta-W alloys has also been studied [23-27]. I. Ipatova *et al.* [27] have studied the formation and evolution of vacancies in Ta and Ta-W alloys induced by radiation combined with proton irradiation experiments. Random distribution of voids was observed in the microstructure of Ta induced by radiation, and the further increase of damage degree would lead to the continuous nucleation and growth of voids. Compared with Ta (0.1dpa), there were vacancies in W-doped materials with higher damage degree (1.55dpa). The experiment showed that the formation of vacancies under irradiation would be delayed when alloying element W was used as in Ta system. To explore the precipitation mechanism of alloying elements and improve the material radiation resistance, it is important to understand the interaction between alloying elements and radiation point defects. However, the generation of defects in Ta-W alloys and the interaction mechanisms between alloying elements and irradiation defects have been rarely studied.

As with the first wall materials, alloying elements would be precipitated in fusion irradiation environment. The precipitation of alloying elements is closely related to the production of irradiation defects. Experimental investigation would be a long-term process on data accumulation and analysis, and it is difficult to trace the evolution behavior of materials at atomic scale. Therefore, theoretical simulations are particularly important in order to establish the physical pictures of radiation damage in materials. In this paper, first-principles calculations were carried out to simulate the behavior of W and their interaction with vacancy defects in Ta-W alloys. There is no aggregation for W in Ta-W system, even the vacancy would not become the nucleation center of W atoms. Besides, the mechanism of the inhibition of vacancy defects by W in Ta system was also explained. Our calculations provide a theoretical basis for further studies on the radiation resistance of Ta and related alloys.

## 2. Computational details

First-principles density-functional-theory calculations, as implemented in VASP [28,29], are performed to relax atomic structures and investigate electronic properties. The Perdew-Burke-Ernzerhof (PBE) exchange-correlation functional [30] and the projector-augmented-wave potentials [31] are used. A 4×4×4 supercell of body-centered cubic Ta including 128 atoms is adopted in our simulations. An energy cutoff of 350 eV for the plane-wave basis and a 3×3×3 Monkhorst-Pack k-mesh [32] are adopted. Atomic coordinates are optimized with a convergence threshold of $1\times10^{-5}$ eV/Å on interatomic forces. The relaxed parameter of the supercell is 3.307 Å, which is in good agreement with the experimental value of 3.306 Å [33].

## 3. Results and discussions

### 3.1. The aggregation behavior of W in Ta

The binding energies $(E_b)$ of W in Ta system were calculated in order to reflect the system stability, described as

$$E_b = nE_{sol} - E_{nsol} - (n-1)E_{per} \qquad (1)$$

where $n$ is the number of W atoms in the Ta-W alloy system, $E_{sol}$ is the energy of a single W atom in the system and $E_{nsol}$ is the energy of $n$ W atoms in the system. A positive value of $E_b$ represents an attractive interaction between W atoms, while a negative value represents repulsive interaction. The W-W binding energies with 128 atoms have been calculated.

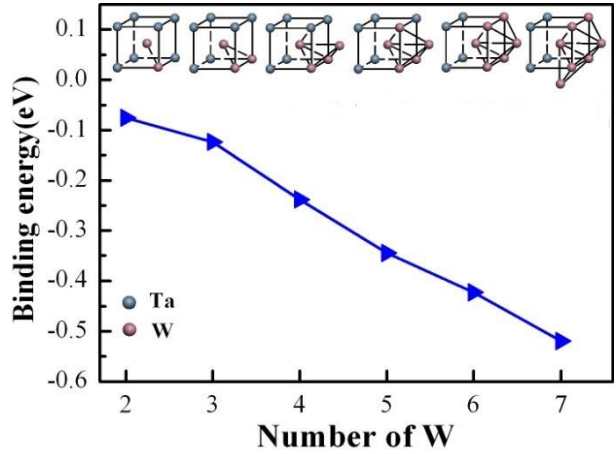

Figure 1. W-W binding energies in Ta-W alloy system. The illustrations are the cluster configurations of W in Ta system.

The calculated W cluster binding energies are plotted in figure 1. It can be clearly seen from the diagram that the W-W binding energies are negative and decrease gradually with the increase of the number of W atoms. The repulsive $E_b$ indicates that W is not likely to aggregate but is more inclined to be dispersed. The supercell size has a slight effect on the binding energy.

### 3.2. Stability of mono-vacancy in Ta-W system

In this section, we first consider the effect of mono-vacancy on W aggregation in Ta-W system. For a pure Ta lattice, the formation energy of mono-vacancy ($E_{vac}^{f}$) is described as

$$E_{vac}^{f} = E_{vac} - \left(\frac{N-1}{N}\right)E_{per} \qquad (2)$$

where $E_{vac}$ is the total energy of the supercell with a mono-vacancy, $E_{per}$ is the total energy of the pure Ta supercell with N atoms. The formation energy of a mono-vacancy will be given by

$$E_{vac}^{f} = E_{vac}^{Ta-W} - E_{per}^{Ta-W} + \frac{1}{N} E_{per} \quad (3)$$

where $E_{vac}^{Ta-W}$ is the total energy of the Ta-W supercell with a mono-vacancy, and $E_{per}^{Ta-W}$ is the total energy of the supercell contains $N$-1 Ta atoms and one W atom. The mono-vacancy is placed in nearest neighbor (NN) locations as 1NN, 2NN, 3NN, 4NN, respectively, according to the position of solute atoms and the symmetry of crystal structure.

We denote $E_{W-V}$ as binding energy between the W atom and mono-vacancy. As shown in figure 2, $E_{W-V}$ first changes gradually from negative to positive and then decreases to be negative again, with the increase of the distance between W and mono-vacancy. The binding energy of mono-vacancy at 1NN position is the minimum ($E_{W-V} = -0.298 eV$) and that of mono-vacancy at 2NN is the maximum ($E_{W-V} = 0.098 eV$). It is shown that the interactions between W and mono-vacancy are generally repulsive, except that for the 2NN site. Therefore, the mono-vacancy could only exist stably with the W atom when it locates in 2NN sites.

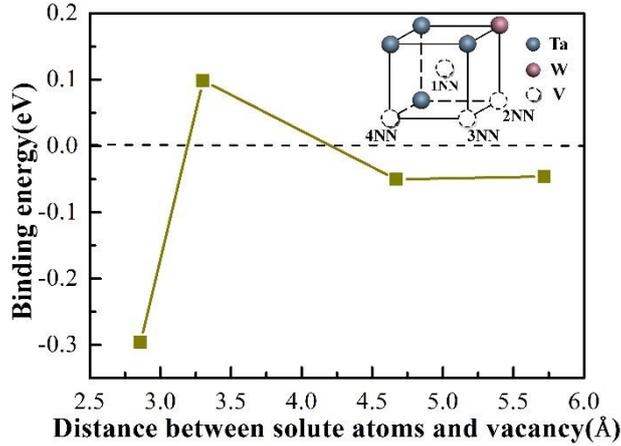

Figure 2. The binding energy $E_{W-V}$ between W and mono-vacancy at different positions in Ta-W system, the schematic diagram represents the mono-vacancy model in the Ta-W system, where 1NN, 2NN, 3NN, 4NN are the four nearest neighbors around the W atom, and V is the mono-vacancy.

Figure 3 shows the formation energy $E_{vac}^{f}$ of mono-vacancy at different positions. $E_{vac}^{f}$ is at its maximum at the 1NN site (3.139 eV), and decreases to its minimum at the 2NN site (2.746 eV). With the increase of the W-vacancy distance to the 3NN and 4NN sites, $E_{vac}^{f}$ becomes closer and closer to its value in pure Ta (2.844 eV), which is reasonable since the interaction between W and vacancy will become weaker and weaker. Compared with pure Ta, the existence of W atom is beneficial to the formation of mono-vacancy only when appearing at the 2NN site, since the only case that $E_{vac}^{f}$ becomes smaller than that in pure Ta.

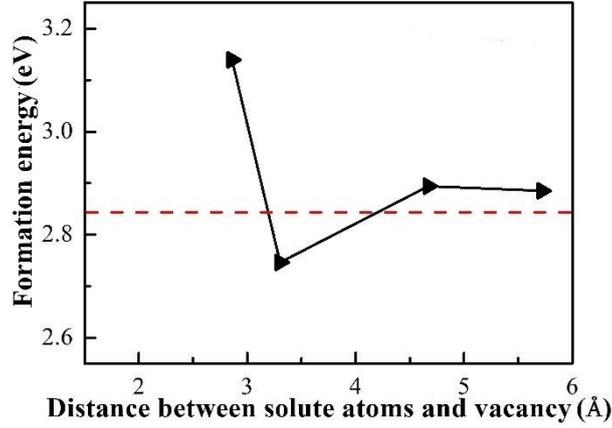

Figure 3. The formation energies $E_{vac}^f$ of mono-vacancy at different nearest neighborhood positions in Ta-W system. The dash line represents the formation energy in pure Ta.

## 3.3. Effect of vacancies on W aggregation in Ta-W system

When vacancies are present in the Ta-W system, the binding energy between vacancies and solute atoms can be given by:

$$E_b(vac_m sol_n) = mE_{vac} + nE_{sol} - E_{vac_m sol_n} - (m + n - 1)E_{per} \qquad (4)$$

Where *m* is the number of vacancies in Ta-W alloy, $E_{vac}$ is the total energy of mono-vacancy in the pure Ta lattice, and $E_{vac_m sol_n}$ is the total energy of Ta-W system containing *m* vacancies and *n* W solute atoms.

The binding energy $E_{W-V}$ between a mono-vacancy and W atom in Ta-W-V system and $E_{W-2V}$ between di-vacancies and W atom in Ta-W-2V system are calculated, respectively, as shown in Fig. 4. With the increase of W concentration, the binding energy $E_{W-V}$ presents a negative value and increases firstly then decreases gradually, which indicates the repulsive interaction between the mono-vacancy and the W atom and correspondingly the aggregation characters of W would not be changed. The binding energy $E_{W-2V}$ shows a negative value and decreases gradually with the increase of W concentration in Ta-W-2V system, indicating even the di-vacancies would not promote the aggregation of W. Compared with the mono-vacancy, the di-vacancies enhance the repulsive effect between vacancy and W clusters.

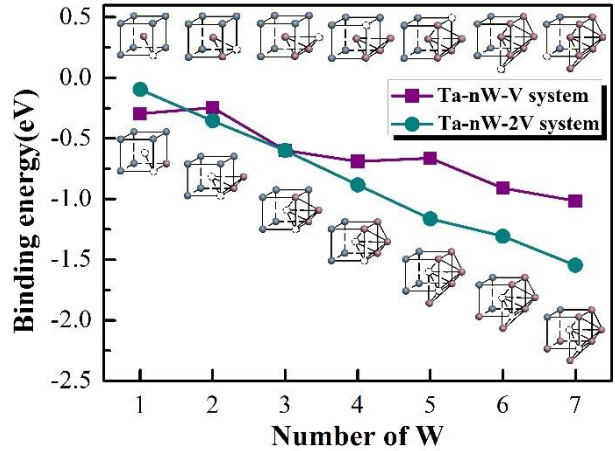

Figure 4. The binding energies between W and vacancy in Ta-W alloys. The illustrations show the Ta-*n*W-V and Ta-*n*W-2V systems, respectively.

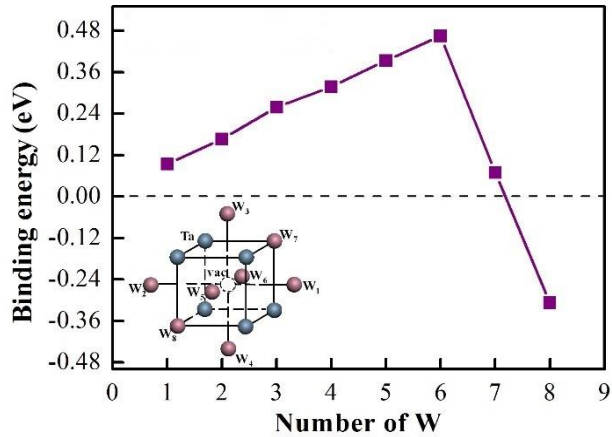

Figure 5. The binding energies between mono-vacancy and W. The illustration represents the Ta-W alloys system containing mono-vacancy, where all the W atoms locate in the 2NN position of the mono-vacancy.

As seen from the profile in figure 4, the repulsive interaction between W and vacancy decreases when the number of W atoms is 2.

Since the vacancy is at the 2NN position, the W atom and the vacancy appear to be attractive and could co-exist stably. As shown in figure 5, when the W atoms are in the 2NN positions of mono-vacancy ($n$=1~6), they could be stable around the vacancies. However, when the W atoms continue to increase ($n$=7~8), the binding energy drops to negative value. Although vacancies would form clusters in W in a special circumstance, it would eventually lead to a dispersed distribution of W clusters as the number of W or vacancy increases. Therefore, the existence of W also hinders the aggregation of vacancy.

### 3.4. Effect of W on vacancy aggregation in Ta-W system

Vacancies tend to form clusters as temperature increases in material under irradiation. We mainly investigated the interaction between solute W and vacancy clusters in Ta-W alloys. The binding energies of vacancy clusters and W atoms ($E_{vac_m sol_n}$) in different configurations are calculated by Eq. (4), as shown in figure 6.

The binding energies $E_{vac_m sol_n}$ in Ta-W-Vac system and Ta-2W-Vac system appear to be negative at first with the increase of vacancy number, then become positive and continuously increase. With the increase of W atoms, the interaction among vacancies changes from repulsion to attraction. It turns out that the stability of vacancy clusters is related to both the number W and vacancy. The $E_{vac_m sol_1}$ with vacancy number 2 is negative ($-0.295 eV$), and the corresponding value in Ta-2W-Vac system is $-0.353 eV$, indicating that the vacancies become more dispersed because of the increase of W. When the vacancy number is 3, the $E_{vac_m sol_1}$ in the Ta-W-Vac system becomes positive, but it is still negative in the Ta-2W-Vac system. When the vacancy number is larger than 3, $E_{vac_m sol_n}$ are all positive and increase gradually. The results show the vacancies present a varied process from an initial repulsive interaction to the final aggregation when solute W induced into Ta lattice. Such vacancy behavior can be understood in this way, the vacancy in Ta-W can be regarded as two categories: "vacancy-vacancy" and "W-vacancy" system. The interaction between "W-vacancy" is dominated when there are few vacancies in the system, therefore the repulsive interaction would be enhanced with the increase of the W atoms. However, with the increase of vacancies, the initial attraction between vacancies would increase continuously and consequently lead to the nucleation and slowly growing up.

Figure 6 shows that the minimum nucleation number of vacancy is affected by W, increasing from 2 to 3 in Ta-2W-Vac system. It indicates that the increase of W concentration could inhibit vacancy nucleation. Our calculation is in agreement with experimental results, where the void density decreased due to the addition of W in Ta under irradiation [27]. With the increase of irradiation time, there will be more and more vacancies and will eventually lead to the formation of voids. However, the addition of W could inhibit the damage process to some extent.

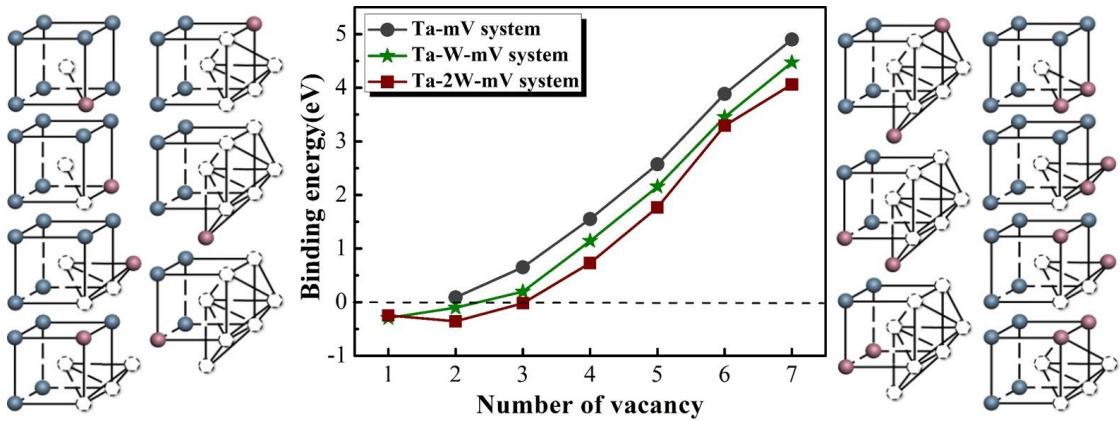

Figure 6. Binding energies of vacancies with different number of solute W atoms, where W is placed in the 1NN and 2NN position of the vacancy clusters, respectively. The left (right)

illustrations represent the Ta-W-*m*V (Ta-2W-*m*V) configurations.

## 4. Conclusions

In this paper, the interaction between W solute atoms and vacancies in Ta-W alloys have been calculated and analyzed by first-principles calculations, based on density functional theory. W atoms are more likely to distribute dispersedly than form clusters. W is favorable to the produce of vacancy, but hinders the aggregation of vacancy, especially with the increase of W atoms. We calculated the interaction between W and vacancies in Ta-W-Vac and Ta-2W-Vac systems, showing that the binding energies between W and vacancies are affected by their concentrations. The increase of vacancy concentration would gradually offset the repulsion between W and vacancies, leading to the nucleation and growth of the vacancy cluster. While when the concentration of W increases, the repulsive interaction between W and vacancies would be enhanced, which can hinder the process of vacancy aggregation. Therefore, as an alloying element, W can suppress the aggregation of vacancy defects in Ta to a certain extent, and thus can be doped into Ta-based materials as a point defect repair element. Our calculations provide a theoretical explanation for the delay of the evolution of vacancies as well as voids through the solute effect in Ta-W alloys in damage irradiation experiments, and could be supplied as a theoretical foundation to improve the radiation resistance of Ta-based materials.

## References

[1] Xu A, Armstrong D E J, Beck C, Moody M P, Smith G D W, Bagot P A J and Roberts S G 2017 *Acta. Mater.* 124 71–8
[2] Novakowski T J, Tripathi J K, and Hassanein A 2016 *Sci. Rep.* 6 39746
[3] Kaufmann D, Mönig R, Volkert C A and Kraft O 2011 *Int. J. Plasticity.* 27 470–8
[4] Zhou J, Akhtar S K, Cai R and Chen L 2006 *J. Iron. Steel. Res. Int.* 13 68–74
[5] Shimko D A, Shimko V F, Sander E A, Dickson K F and Nauman E A 2005 *Wiley Periodicals, Inc.* 73B 315–24
[6] Zardiackas L D, Parsell D E, Dillon L D, Mitchell D.W, Nunnery L A and Poggie R 2001 *John Wiley & Sons, Inc.* 58 181–7
[7] Deng C, Liu S F, Ji J L, Hao X B, Zhang Z Q and Liu Q 2014 *J. Mater. Process. Tech.* 214 462–9
[8] Abedin S Z E, Welz-Biermann U and Endres F 2005 *Electrochem. Commun.* 7 941–6
[9] Schuster B E, Ligda J P, Pan Z L and Wei Q 2011 *JOM.* 63 27–31
[10] Buckman R W 2000 *JOM* 52 40–1
[11] Silva R A and Guidoin R 2002 *J. Mater. Sci:Mater. M.* 13 495–500
[12] Souza K A and Robin A 2007 *Mater Chem Phys.* 103 351–60
[13] Robin A and Rosa J L 2000 *Int. J. Refract, Met. H.* 18 13–21
[14] Lee Y J, Lee T H, Kim D Y, Nersisyan H H, Han M H, Kang K S, Bae K K, Shin Y J and Lee J H 2013 *Surf. Coat. Tech.* 235 819–26
[15] Novakowski T J, Sundaram A, Tripathi J K, Gonderman S and Hassanein A 2018 *J. Nucl.*

*Mater.* 504 1–7

[16] Cui J, Zhao L, Zhu W, Wang B, Zhao C, Fang L and Ren F 2017 *J. Mech. Behav, Biomed.* 74 315–23

[17] Lin Z, Lavernia E J and Mohamed F A 1999 *Acta Mater.* 47 1181–94

[18] Nemat-Nasser S and Kapoor R 2001 *Int. J. Plast.* 17 1351–66

[19] Perez-Prado M T, Hines J A and Vecchio K S 2001 *Acta Mater.* 49 2905–17

[20] Gao F, Zhang X, Ahmad S, Wang M, Li L, Xiong W and Zhang J 2017 *Rare Metal. Mat. Eng.* 46 2753–62

[21] Browning P N, Alagic S, Carroll B, Kulkarni A, Matsond L and Singh J 2017 *Mater. Sci. & Eng. A* 680 141–51

[22] Nemat-Nasser S and Isaacs J B 1997 *Acta Mater.* 45 907–19

[23] Zhang J, Ma G Q, Godfrey A, Shu D Y, Chen Q and Wu G L 2017 *Mater. Sci. Eng.* 219 012051

[24] Wang S, Niu L, Chen C, Jia Y L, Wang M P, Li Z, Zhong Z H, Lu P, Li P and Wu Y C 2017 *J. Alloy. Compd.* 699 57–67

[25] Chen C, Wang S, Jia Y L, Wang M P, Li Z, Zhong Z H, Lu P, Wu Y C and Cao L F 2016 *Int. Journal of Refractory Metals & Hard Mater.* 61 136–46

[26] Wang S, Chen C, Jia Y L, Wang M P, Li Z and Wu Y C 2016 *Int. Journal of Refractory Metals & Hard Mater.* 54 104–15

[27] Ipatova I, Wady P T, Shubeita S M, Barcellini C, Impagnatiello A and Jimenez-Melero E 2017 *J. Nucl. Mater.* 495 343–50

[28] Kresse G and Hafner J 1993 *Phys. Rev. B* 47 558–61

[29] Kresse G and Furthmüller J 1996 *Phys. Rev. B* 54 11169–86

[30] Perdew J P, Burke K and Ernzerhof M 1996 *Phys. Rev. Lett.* 77 3865–8

[31] Blöchl P E 1994 *Phys. Rev. B* 50 17953–79

[32] Monkhorst H J and Pack J D 1976 *Phys. Rev. B* 13 5188–92

[33] Nestell J J E and Christy R W 1979 *Phys. Rev. B* 21 3173–9